\documentclass{article}
\usepackage{ijcai23}

\usepackage{times}
\usepackage{soul}
\usepackage{url}
\usepackage[hidelinks]{hyperref}
\usepackage[utf8]{inputenc}
\usepackage[small]{caption}
\usepackage{graphicx}
\usepackage{amsmath}
\usepackage{amsthm}
\usepackage{booktabs}
\usepackage{algorithm}
\usepackage{algorithmic}
\usepackage[switch]{lineno}
\usepackage{multirow}
\usepackage{comment}
\usepackage{balance}

\usepackage[symbol]{footmisc}

\title{Hybrid Multi-Criteria Preference Ranking by Subsorting{\footnote{Accepted by M-PREF Workshop at IJCAI 2023\\ 
\hspace*{0.6cm}{\textcopyright} 2023 Copyright held by the owner/author(s).}}}

\author{
Yong Zheng$^1$
\and
David Xuejun Wang$^2$
\affiliations
$^1$Illinois Institute of Technology, USA\\
$^2$Morningstar, Inc., USA
\emails
yzheng66@iit.edu, david.wang@morningstar.com 
}

\begin{document}

\maketitle

\begin{abstract}
Multi-criteria recommender systems can improve the quality of recommendations by considering user preferences on multiple criteria. One promising approach proposed recently is multi-criteria ranking, which uses Pareto ranking to assign a ranking score based on the dominance relationship between predicted ratings across criteria. However, applying Pareto ranking to all criteria may result in non-differentiable ranking scores. To alleviate this issue, we proposed a hybrid multi-criteria ranking method by using subsorting. More specifically, we utilize one ranking method as the major sorting approach, while we apply another preference ordering method as subsorting. Our experimental results on the OpenTable and Yahoo!Movies data present the advantages of this hybrid ranking approach. In addition, the experiments also reveal more insights about the sustainability of the multi-criteria ranking for top-$N$ item recommendations.
\end{abstract}

\section{Introduction}
\noindent

Recommender systems (RS) aim to offer personalized suggestions to users by considering their individual preferences. These preferences can be based on different factors or standards related to the recommended items. Multi-criteria recommender systems (MCRS) have been developed to enhance recommendations in travel/tourism domain (e.g., hotel booking at TripAdvisor.com, restaurant reservations at OpenTable.com)~\cite{borras2014intelligent,zheng2012differential}, and movie domain (e.g., Yahoo!Movies)~\cite{goyani2020review}. Unlike traditional RS, MCRS take user preferences across multiple criteria into consideration, thereby improving the quality of recommendations.

\begin{table}[ht!]\small
\caption{Example of Rating Matrix from OpenTable.com}\label{fig:dataexample}
\begin{tabular}{|c|c|c|c|c|c|c|}
\hline
\textbf{User} & \textbf{Item} & \textbf{Rating} & \textbf{Food} & \textbf{Service} &\textbf{Ambience} & \textbf{Value} \\ \hline
$U_1$           & $T_3$            & 4              & 4             & 3         & 4        & 4 \\ \hline
$U_2$            & $T_2$            & 3               & 3             & 3       & 3          & 3 \\ \hline
$U_3$            & $T_1$            & ?               & ?             & ?     & ?             & ?                \\ \hline
\end{tabular}
\end{table}

Table~\ref{fig:dataexample} presents an example of multi-criteria rating data from OpenTable, where we have users' multi-criteria ratings (i.e., ratings on food, service, ambience and value) on the items in addition to the overall ratings. As a result, the rating prediction task in MCRS refers to predicting the overall rating given by a user on an item, e.g., predicting how $U_3$ rates item $T_1$ without the knowledge of associated multi-criteria ratings, as shown in Table~\ref{fig:dataexample}. 

Most MCRS algorithms~\cite{zheng2023multi} were developed by following a two-stage workflow -- \textit{multi-criteria rating predictions}, where the model predicts how a user rates the item from the perspective of multiple criteria, and \textit{rating aggregations}, where the model estimates the overall rating by a utility function (e.g., linear or non-linear transformation) from the predicted multi-criteria ratings. We recently proposed \textit{multi-criteria ranking} (MCRanking)~\cite{zheng2022multi} which involves a unique utility function that draws from the concept of dominance relation in multi-objective optimization (MOO). MCRanking derives a ranking score for each item candidate based on the predicted multi-criteria ratings of the item. Pareto ranking methods were initially utilized to infer the ranking score which refers to the number of other item candidates that a given item can dominate. MCRanking was demonstrated as one of the most effective MCRS algorithm in comparison with the state-of-the-art MCRS recommendation approaches.


However, MCRanking faces certain challenges relating to stability and dimensionality, if we apply Pareto ranking to all criteria. More specifically, it is possible for multiple items to dominate the same number of other items, resulting in identical rankings. Since Pareto ranking cannot differentiate between these items, this issue can lead to instability. Furthermore, as the number of criteria increases, it becomes more likely for items to share the same rank, exacerbating the dimensionality issue. In the area of MOO, the similar issue also occurs. Researchers proposed to utilize other preference ordering methods, such as the \textit{relaxed Pareto ranking methods}~\cite{farina2004fuzzy,laumanns2002combining} and \textit{regular preference ordering approaches}~\cite{bentley1998finding,kukkonen2007ranking,garza2009ranking} which are not necessary to satisfy the principles of Pareto dominance.

Our previous work have tried these regular preference ordering~\cite{zheng2023ordering} and relaxed Pareto ranking methods~\cite{zheng2023relaxed}, respectively. However, using each of these ranking methods individually can still result in non-differential ranking scores. In this paper, we propose and exploit a hybrid multi-criteria ranking method, where two ranking methods can be fused together by using one of them as major sorting and another one as subsorting. Our experimental results on the OpenTable and Yahoo!Movies data present the advantages of this hybrid ranking approach. In addition, the experiments also reveal more insights about the sustainability of the multi-criteria ranking for top-$N$ item recommendations.

\section{Related Work}
\subsection{Multi-Criteria Recommendations}
In MCRS, we have user preferences on multiple criteria in addition to users' overall rating on the items. Therefore, the utility function of how a user likes an item in MCRS is no longer a function only with the overall rating, but also the multi-criteria ratings which can represent complex user tastes. More specifically, traditional RSs only take users' overall ratings on the items into consideration, as shown by Equation~\ref{eq:rs}. $R_0$ is the overall rating given by a user on one item.
\begin{equation}\label{eq:rs}
R: Users \times Items \rightarrow R_0
\end{equation}

MCRS additionally consider user's ratings on different aspects of the items, as shown by Equation~\ref{eq:mcrs}. We assume there are $k$ aspects or criteria of the items, and users may give ratings to each aspect in addition to the overall rating $R_0$.
\begin{equation}\label{eq:mcrs}
R: Users \times Items \rightarrow R_0 \times R_1 \times R_2 \times ... \times R_k
\end{equation}

The existing MCRS algorithms can be classified into two categories. The first category follows a two-stage workflow, as described below.
\begin{enumerate}
\item \textit{multi-criteria rating predictions}, where the model predicts how a user rates the item from the perspective of multiple criteria through independent~\cite{adomavicius2007new} or dependent approaches~\cite{zheng2017criteria,nassar2020multi}
\item \textit{rating aggregations}, where the model estimates the overall rating by aggregation from the predicted multi-criteria ratings. The aggregation function is also known as the multi-attribute utility function in multi-criteria decision making~\cite{san2012multi}. Most algorithms in this category focused on the development of utility functions. These functions utilized in MCRS include but not limited to -- linear~\cite{adomavicius2007new} and non-linear predictive models, such as support vector regression~\cite{fan2013robust}, distance-based utility function~\cite{zheng2019utility,zheng2020penalty}, multi-layer perceptron model~\cite{nassar2020multi}. 
\end{enumerate}

From the perspective of multi-criteria decision making (MCDM)~\cite{doumpos2019preference,zheng2023multi}, most MCDM methods, such as multi-attribute utility functions, outranking, multi-objective programming and preference disaggregation models, can be utilized to solve the problems in MCRS. However, recommender systems need a specific score to sort items in order to produce top-$N$ recommendations. As a result, the utility functions became the most popular MCDM methods to be reused in MCRS, especially in the stage of rating aggregations as indicated above.

The second category of the models can estimate the overall rating directly, without predicting multi-criteria ratings in advance, e.g., the heuristic method~\cite{adomavicius2007new} just utilized multi-criteria ratings to find better user neighborhoods to help predict the overall rating by collaborative filtering. Hong et al. proposed to utilize tensor factorization~\cite{hong2021multi}, where ratings in different criteria and the overall ratings were just considered as contextual ratings in the tensor. Namely, they considered the scenario of rating (e.g., in overall scenario or each criterion) as contexts. It is different from our previous work~\cite{zheng2017situation,zheng2019multi} in which we considered the predicted rating in each criterion as contexts.

\subsection{Preliminary: Multi-Criteria Ranking}
The concept of MCRanking~\cite{zheng2022multi} is based on a universal principle, which involves inferring a ranking score directly from the predicted multi-criteria ratings, rather than estimating the overall ratings and subsequently sorting items in MCRS. Pareto ranking is a ranking method developed in the field of multi-objective evolutionary algorithms (MOEAs). It relies on the definition of dominance relation. We adapted the the Fonseca and Fleming's ranking~\cite{fonseca1993genetic} as the Pareto ranking method to MCRanking and defined the dominance relation as shown by Equation~\ref{eq:cond1} and~\ref{eq:cond2}. 
\begin{equation}\label{eq:cond1}
 R_{u, T_i, m}\geq R_{u, T_j, m}, \forall m\in\{1..M\}
\end{equation}

\begin{equation}\label{eq:cond2}
  R_{u, T_i, m}> R_{u, T_j, m}, \exists m\in\{1..M\}
 \end{equation}

Assume there are $M$ criteria to be considered, we use $R_{u, T_i, m}$ to represent user $u$'s rating on the item $T_i$ with respect to the $m^{th}$ criterion. The item $T_i$ dominates $T_j$, if and only if they satisfy the following conditions.

\begin{itemize}
	\item The multi-criteria ratings given by user $u$ on item $T_i$ must be no less than the corresponding ratings given to the item $T_j$, as shown by Equation~\ref{eq:cond1}.
	\item There is at least one criterion where the associated rating given to $T_i$ is higher than the one given to $T_j$, as described by Equation~\ref{eq:cond2}. 
\end{itemize}

	The ranking score, therefore, equals to the number of other items that a given item can dominate. Note that we did not transform the MCRS to a problem of multi-objective recommendations~\cite{zheng2022survey}. We just took advantage of the notion of dominance relation to derive the ranking score for MCRanking.

Take the data in Table~\ref{tab:prank} for example, there are five item candidates to be ranked for a specific user, and we already obtained the predicted multi-criteria ratings associated with each item. $T_1$ is the best item since it can dominate all other candidates. $T_3$ is the worst since it cannot dominate any other items. $T_2$ and $T_5$ are at the same rank since they can dominate two items, but there is no dominance relation between $T_2$ and $T_5$. The ranking score derived from Pareto ranking can be shown by the last column in Table~\ref{tab:prank}. These scores can be used to produce top-$N$ item recommendations.

\begin{table}[ht!]\small
    \centering
 \caption{Example of Pareto Ranking}\label{tab:prank}
    \begin{tabular}{|c|c|c|c|c|c|}
\hline
 \textbf{Item} & \textbf{Rating} & \textbf{Food} & \textbf{Service} & \textbf{Ambience} & \textbf{Ranking Score}\\ \hline
 T$_1$             & ?               & 5             & 5                & 5    & 4          \\ \hline
 T$_2$             & ?               & 4             & 4                & 4       &  2       \\ \hline
T$_3$             & ?               & 3             & 3                & 3     &  0         \\ \hline
 T$_4$             & ?               & 4             & 3                & 3       &  1       \\ \hline
 T$_5$             & ?               & 4             & 5                & 3     & 2         \\ \hline
\end{tabular}
\end{table}

There are two potential issues when we use Pareto ranking for MCRanking. First, Pareto ranking cannot distinguish items with same ranks, e.g., $T_2$ and $T_5$ have a same rank. These items may be given random sequence in the recommendation list, e.g., $T_2$ ranked higher than $T_5$, or vice versa. It further results in instability in performance. Moreover, there are more items that share a same rank, if the number of criteria increases, which is referred as the dimensionality issue. 

There are two potential solutions to alleviate the issues above. First, there are similar issues in the area of MOO, and researchers have developed relaxed Pareto ranking methods to alleviate the issues. In addition, we can try to reduce the number of criteria dimensions to be involved in Pareto ranking. And even more, we can assign personalized criteria dimensions in Pareto ranking, i.e., we can use different criteria dimensions for different users in the ranking process.

\subsection{Relationships with MCDM}
\noindent
As mentioned previously, recommender systems need a specific score to sort items in order to produce top-$N$ recommendations. As a result, the utility functions from MCDM theories became the most popular MCDM methods to be reused in MCRS. Other methods, such as outranking, may not be appropriate, since they may produce a set of items which are superior over other items, and they cannot differentiate the ranks of this set of items. Accordingly, the multi-criteria ranking method mentioned above can also be considered as a special utility function, since we are able to obtain a numerical score for each individual item.

\section{Hybrid Multi-Criteria Rankings}
\noindent
In this section, we introduce selected relaxed Pareto ranking and regular preference ordering techniques first, followed by the discussion of our proposed hybrid multi-criteria rankings.

\subsection{Relaxed Pareto Ranking by $k$-dominance}
There are two major relaxed Pareto ranking methods -- the $\epsilon$-dominance~\cite{laumanns2002combining} and the $k$-dominance method~\cite{farina2004fuzzy}. The $\epsilon$-dominance is straightforward, where a disturbance factor is introduced to fine-tune the models. Our previous work~\cite{zheng2023relaxed} revealed that the $k$-dominance method~\cite{farina2004fuzzy} was a better choice for multi-criteria ranking. 

More specifically, the $k$-dominance method introduces the fuzzy definition of optima. In the following discussions, we use $M$ to denote the number of criteria, $u$ as a user, $T_i$ and $T_j$ as two items. First of all, three numbers are defined to distinguish three scenarios:

\begin{equation}\label{eq:nb}
n_b(T_i, T_j) = |\left\{m|R_{u, T_i, m} > R_{u, T_j, m}\right\}|, m\in\{1..M\}
\end{equation}

\begin{equation}\label{eq:ne}
n_e(T_i, T_j) = |\left\{m|R_{u, T_i, m} = R_{u, T_j, m}\right\}|, m\in\{1..M\}
\end{equation}

\begin{equation}\label{eq:nbw}
n_w(T_i, T_j) = |\left\{m|R_{u, T_i, m} < R_{u, T_j, m}\right\}|, m\in\{1..M\}
\end{equation}

Namely, $n_e$ refers to the number of criteria where $T_i$ and $T_j$ received same ratings from user $u$. $n_b$ equals to the number of criteria where $T_i$ has a higher rating than $T_j$ given by $u$. $n_w$, therefore, refers to the number of criteria where $T_i$ has a lower rating than $T_j$ given by $u$. The summation of $n_e$, $n_b$ and $n_w$ equals to the number of criteria, i.e., $M$.

$T_i$ is said to $k$-dominate $T_j$ if and only if Equation~\ref{eq:cond5} and~\ref{eq:cond6} are satisfied.

\begin{equation}\label{eq:cond5}
n_e<M
\end{equation}

\begin{equation}\label{eq:cond6}
n_b\geq\frac{M-n_e}{k+1}=\frac{n_b+n_w}{k+1}
\end{equation}

In other words, Equation~\ref{eq:cond5} tells that there is at least one inequity (i.e., one criterion) in multi-criteria ratings between two items. In addition, the dominance is relaxed by Equation~\ref{eq:cond6}. $k$ is the relaxation factor, where $0 \leq k \leq 1$. It becomes the original Pareto ranking, if $k$ is set to zero.

The $k$-dominance method takes into account the number of criteria where the first
candidate item is better than the second one and vice versa, which results in a more general
definition, being able to cope with a wider variety of pairwise comparisons among two items in the set of candidate items to be recommended.

\subsection{Regular Preference Ordering}
The preference ordering approaches here refer to the methods which utilize multiple criteria or attribute to sort and rank items. We discuss four examples of these preference ordering methods. For consistency with the notations used in these methods, we assume that a lower rank value corresponds to a higher position in the recommendation list. Thus, the top item in the recommendation list has a rank value of one.

The \textit{average ranking} (AR) and \textit{maximum ranking} (MR) are two popular preference ordering methods~\cite{bentley1998finding,kukkonen2007ranking}. The idea behind them is to find ranks of solutions (or items) in terms of each separate objective (or criterion) and use these ranks with a suitable aggregation function to calculate the fitness value (i.e., ranks of the items). These methods derive the final ranking score from individual ranks associated with each single criterion. Therefore, the results by AR and MR are not guaranteed to satisfy Pareto dominance based on multiple criteria.

The AR method can be described by Equation~\ref{eq:ar}, where it is a summation form rather than an averaging score. The $r_m(T_i)$ refers to the rank of item $T_i$ among all candidate items with respect to the $m^{th}$ criterion. The final rank of item $T_i$, therefore, can be calculated as a summation or average value from these individual ranks over $M$ criteria.

\begin{equation}\label{eq:ar}
Rank(T_i) = \sum_{m=1}^M r_m(T_i)
\end{equation}

The MR method uses the best rank among individual ranks from all criteria, as shown by Equation~\ref{eq:mr}.

\begin{equation}\label{eq:mr}
Rank(T_i) = \min_{m=1}^M r_m(T_i)
\end{equation}

In contrast to these simple ranking-dominance methods, Garza-Fabre et al. proposed the following two methods which can take the degree of dominance into considerations~\cite{garza2009ranking}.

The \textit{global detriment} (GD) method derives the rank by accumulating the difference (e.g., difference of ratings) by which it is inferior to every other solution (or item), with respect to each objective or criterion, as shown by Equation~\ref{eq:gd} and~\ref{eq:pg2}. It equals to the summation of non-negative differences of criteria ratings (i.e., gains in Equation~\ref{eq:pg2}) associated with each criterion based on a pairwise comparison between a given item $T_i$ and other candidate items for a user $u$.

\begin{equation}\label{eq:gd}
Rank(T_i) = \sum_{T_i \neq T_j}gain(T_i, T_j)
\end{equation}

\begin{equation}\label{eq:pg2}
gain(T_i, T_j) = \sum_{m=1}^M max(0, R_{u,T_i,m}-R_{u,T_j,m})
\end{equation}

The \textit{profit gain} (PG) method calculates the rank as the profits through a pairwise comparison between a given item $T_i$ and other candidate items for a user $u$, as shown by Equation~\ref{eq:pg1}. 

\begin{equation}\label{eq:pg1}
Rank(T_i) = \max_{T_i \neq T_j}(gain(T_i, T_j)) - \max_{T_i \neq T_j}(gain(T_j,T_i)))
\end{equation}

It is obvious that the AR and MR methods are simple and straightforward, but they do not satisfy Pareto dominance from the perspective of multiple criteria. The GD and PG methods take advantage of the degree of dominance, rather than defining the ranking score by dominance relation directly. Our previous experiments on MCRS~\cite{zheng2023ordering} found that GD and PG methods could outperform others in most cases. AR or MR may also deliver good results in specific data sets.

\subsection{Hybrid Multi-Criteria Rankings}
The major problem we would like to alleviate is the non-differential ranking produced by the Pareto ranking methods. Take Table~\ref{tab:top10} for example, there are top-10 items for a specific user. If the system is going to produce top-5 item recommendations for the user, the first two items are included, where we still need three items to be selected from the items with the same ranking score (i.e., 4). This is a common issue in Pareto ranking and even the relaxed Pareto ranking approaches, since multiple items may have a same rankings score due to the definition of the dominance relation over multiple criteria.

\begin{table}[ht!]\small
\centering
\caption{Top-10 Items}\label{tab:top10}
\begin{tabular}{|c|c|}
\hline
\textbf{ItemID} & \textbf{Ranking Score} \\ \hline
22              & 5                      \\ \hline
14              & 5                      \\ \hline
5               & 4                      \\ \hline
46              & 4                      \\ \hline
13              & 4                      \\ \hline
8               & 4                      \\ \hline
65              & 4                      \\ \hline
27              & 4                      \\ \hline
33              & 4                      \\ \hline
55              & 4                      \\ \hline
\end{tabular}
\end{table}

Our proposed hybrid multi-criteria ranking is simple and straightforward. The final hybrid score for ranking can be described by Equation~\ref{eq:hybrid}.

\begin{equation}\label{eq:hybrid}
HybridScore=Score_{major}+Score_{sub}
\end{equation}

We can employ the Pareto ranking or relaxed Pareto ranking as the major sorting method which can produce an integer ranking score (i.e., the number of candidate items an item can dominate). In addition, we use another preference ordering method as subsorting, where the resulting ranking score by subsorting can be normalized to [0. 1) and then added to the score by major sorting, as shown by Equation~\ref{eq:hybrid}. By this way, we can provide decimals to the integer ranking scores. Take Table~\ref{tab:top10} for example, the items with score 4 will be assigned the decimal scores by using subsorting.

In our following experiments, we used k-dominance as the major ranking since our previous work demonstrated its effectiveness over the original Pareto ranking method, and tried to incorporate AR, MR, GD and PG as the subsorting methods.

\section{Experimental Evaluations \& Results}

\subsection{Experimental Setting}
\noindent
In this section, we discuss the results over the OpenTable~\cite{zheng2023open}                                                                 and Yahoo!Movie~\cite{jannach2014leveraging} data sets, as shown by Table~\ref{tab:data}. All ratings in these data were normalized to the scale 1 to 5, and we use 3 as the threshold to determine relevance, i.e., only items with ratings no less than 3 are considered as relevant items in the process of evaluations.

\begin{table}[ht!]\small
\centering
\caption{Data Sets}\label{tab:data}
\begin{tabular}{|c|c|c|}
\hline
                & OpenTable & \begin{tabular}[c]{@{}c@{}}Yahoo!\\ Movie\end{tabular}                       \\ \hline
\# of users                                                           & 1,309                                                                     & 2,162                                                                       \\ \hline
\# of items                                                           & 91                                                                        & 3,078                                                                       \\ \hline
\# of ratings                                                      & 19,537                                                                    & 62,739                                                                      \\ \hline
Criteria       & \begin{tabular}[c]{@{}c@{}}Food\\ Service\\ Ambience\\ Value\end{tabular} & \begin{tabular}[c]{@{}c@{}}Story\\ Direction\\ Acting\\ Visual\end{tabular} \\ \hline
\end{tabular}
\end{table}

\begin{figure}[ht!]
\centering
\includegraphics[scale=0.2]{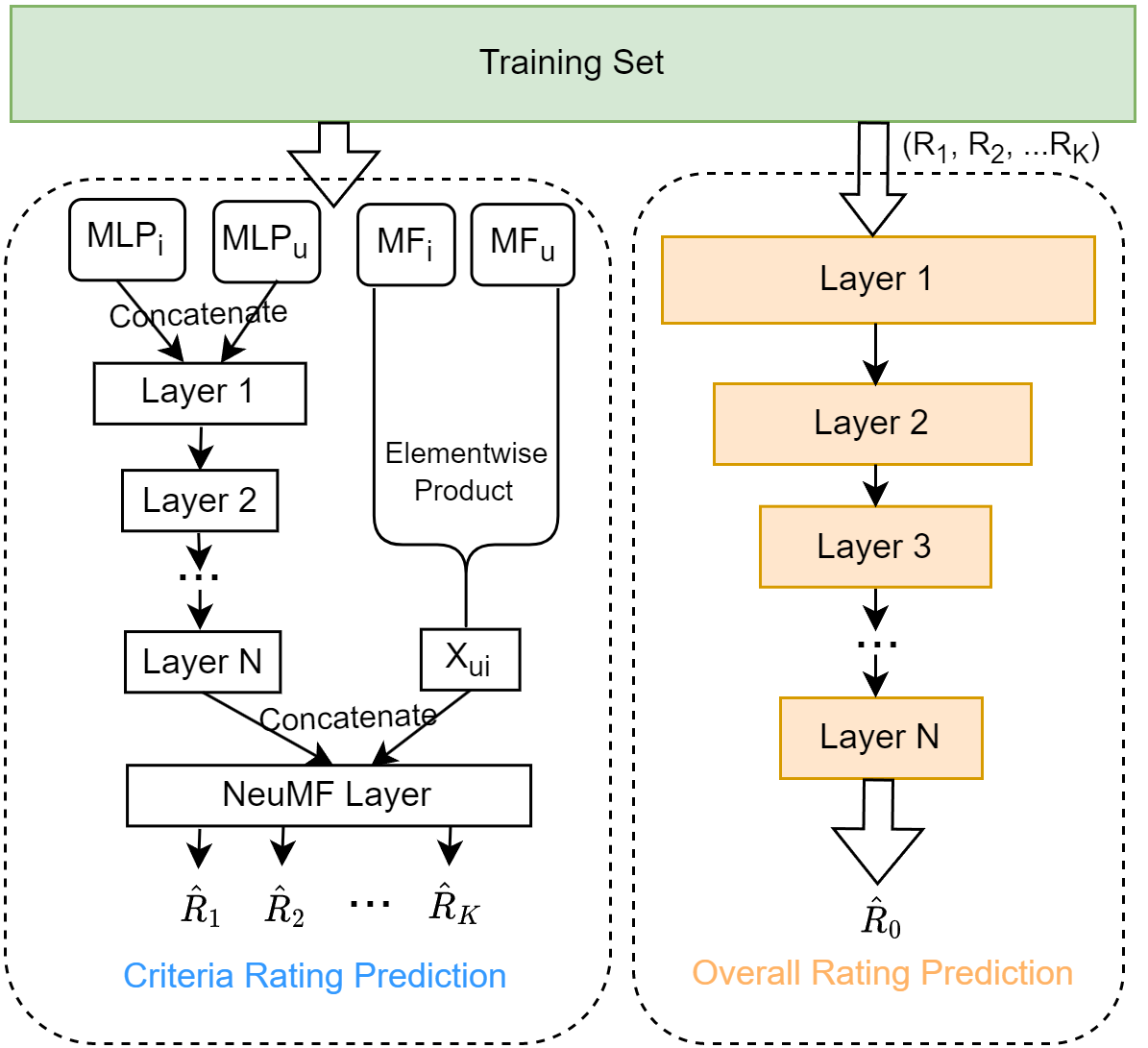}
\caption{The MONeuMF Model}
\label{fig:model}
\end{figure}

\begin{figure*}[ht!]
\centering
\includegraphics[scale=0.75]{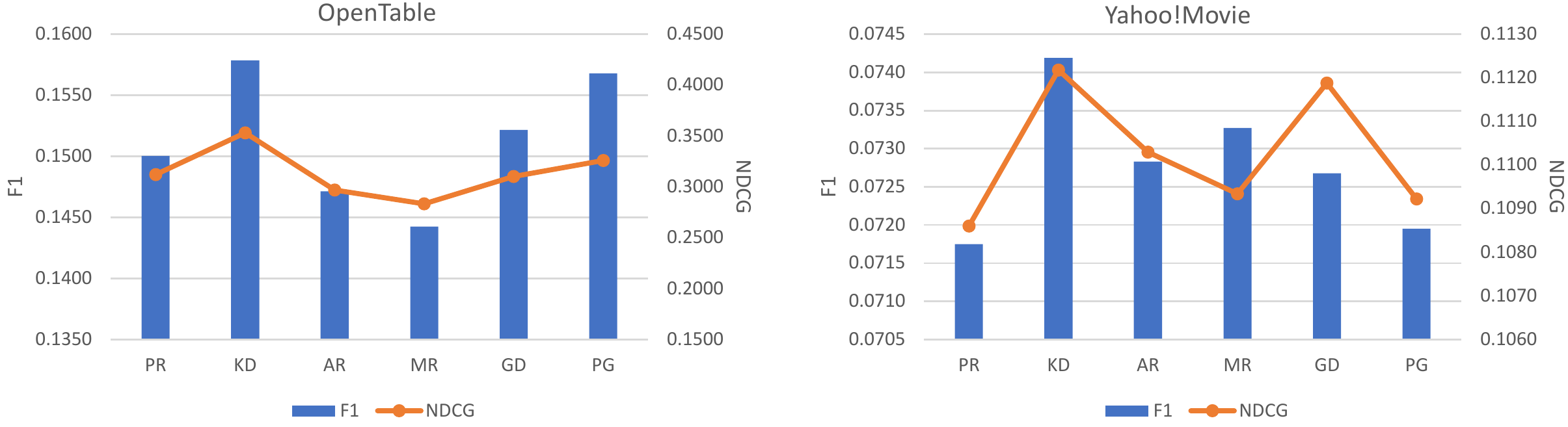}
\caption{Results: Top-10 Item Recommendations By Using Individual Rankings}
\label{fig:rst1}
\end{figure*}
\begin{figure*}[ht!]
\centering
\includegraphics[scale=0.75]{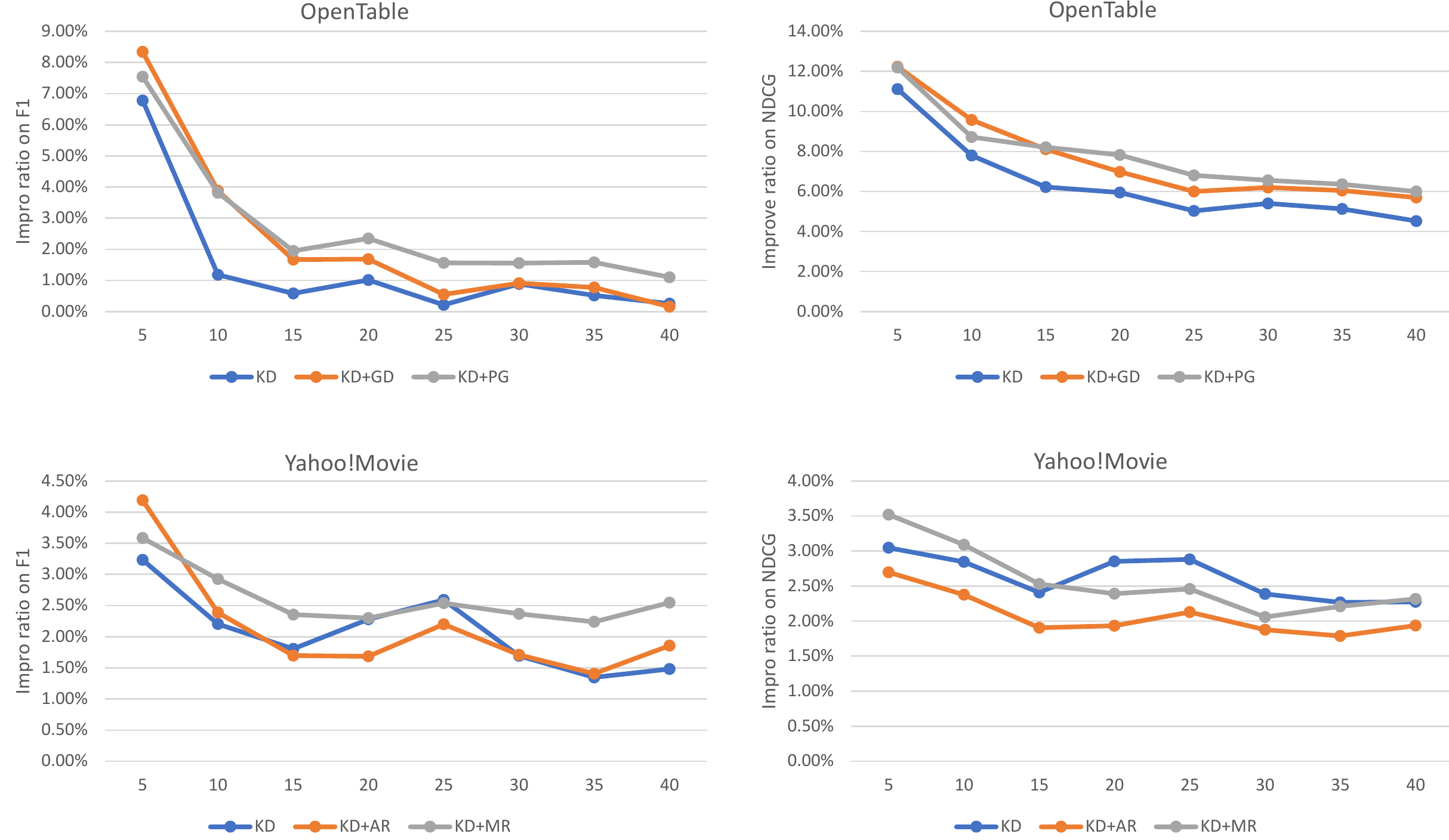}
\caption{Results: Top-N Item Recommendations By Using Hybrid Rankings}
\label{fig:rst2}
\end{figure*}

The multi-output neural matrix factorization (MONeuMF)~\cite{nassar2020multi,zheng2022multi} model, as shown by Figure~\ref{fig:model} is one of most effective MCRS algorithms. It utilizes the neural matrix factorization to predict multi-criteria ratings, where the weighted sum of the rating losses associated with multiple criteria is considered as the joint loss function in the output layer. For simplicity, we use equal weights in experiments. A multi-layer perceptron model is used to estimated the overall rating from multi-criteria ratings.

In our experiments, we use MONeuMF to predict multi-criteria ratings, and then employ different multi-criteria ranking methods based on these predicted multi-criteria ratings, in order to produce top-$N$ item recommendations. Multi-criteria ranking can actually be integrated with any two-stage MCRS algorithms. We selected MONeuMF due to its effectiveness demonstrated in existing research.

First of all, we run top-10 item recommendations over these data sets by using each multi-criteria ranking method individually, including Pareto ranking (PR), the $k$-dominance relaxed ranking (KD), and AR, MR, GD, PG methods. After that, we select KD as the major sorting, and try other preference ordering methods as subsorting for the evaluation purpose. In this step, we also vary the value of $N$ for top-$N$ item recommendations, in order to discover more insights.

Note that we evaluated these models by using 5-fold cross validation on the data sets above and measure the recommendation quality by using the F$_1$ and normalized discounted cumulative gain (NDCG) over the top-$N$ recommendations. 

More specifically, F$_1$ is used to measure the relevance. F$_1$ is a fusion of both precision and recall, as shown by Equation~\mbox{\ref{eq:f1}}. Precision is the fraction of relevant items among the recommended items, while recall is the fraction of relevant items that were retrieved. The number of matched relevant items is usually referred as the number of true positives (TP), while false positives (FP) denote the number of items which are not relevant but shown in the recommendation list, and false negatives (FN) refer to the number of items which are actually relevant but not recommended.

\begin{equation}
Precision=\frac{TP}{TP+FP}
\end{equation}

\begin{equation}
Recall = \frac{TP}{TP+FN}
\end{equation}

\begin{equation}\label{eq:f1}
F_1 =\frac{2\times Precision \times Recall}{Precision+Recall}
\end{equation}

In addition, NDCG is used to examine the ranking quality. NDCG is a metric used for listwise ranking in the well-known learning-to-rank methods. Assuming each user $u$ has a "gain" ${g_{ui}}_j$ from being recommended an item $i$ at rank $j$, the average Discounted Cumulative Gain (DCG) for a list of $J$ items is defined in Equation~\ref{eq:dcg}.

\begin{equation}
\label{eq:dcg}
DCG = \frac{1}{N}\sum_{u=1}^{N}\sum_{j=1}^{J} \frac{{g_{ui}}_j}{max(1,log_2 j))}
\end{equation}

$i_j$ refers to the item at the $j^{th}$ position of the recommendation list. ${g_{ui}}_j$ indicates the gain from this specific item, and it can be calculated from the utility function below, where ${Rel_{i_j}}$ refers to the relevance score (e.g., real rating or rank) of item $i_j$ in the ground truth.

\begin{equation}
\label{eq:ut}
{g_{ui}}_j=2^{Rel_{i_j}}-1
\end{equation}

NDCG, therefore, is the normalized version of DCG given by Equation~\ref{eq:ndcg}, where $DCG^{*}$ is the ideal DCG, i.e., the maximum possible DCG computed from the real ranking of items.

\begin{equation}
\label{eq:ndcg}
NDCG = \frac{DCG}{DCG^{*}}
\end{equation}

\subsection{Results and Findings}
Figure~\ref{fig:rst1} presents the results of top-10 recommendations by using each multi-criteria ranking method, respectively. The bars represent the F$_1$ results with respect to the y-axis on the left, while the curve denotes the NDCG results with respect to the y-axis on the right. In both OpenTable and Yahoo!Movie data sets, the relaxed Pareto ranking method, KD, appears to the optimal approach. The best $k$ in KD is 0.5 and 0.6 for the OpenTable and Yahoo!Movie data, respectively. In terms of the the regular preference ordering approaches, PG is the best one for the OpenTable data, and MR could be the best for the Yahoo!Movie data, though GD presents a higher NDCG.

Afterwards, we examined the performance of the proposed hybrid multi-criteria rankings. In terms of the major sorting, we selected KD, since it delivers the best results as shown by Figure~\ref{fig:rst1}. Note that the KD approach can produce an integer ranking score for each item (i.e., the number of candidate items that an item can dominate). Regarding the subsorting, we tried AR, MR, GD and PG to work together with the KD approach.

Figure~\ref{fig:rst2} presents the results of top-$N$ recommendations by using the hybrid ranking methods. The x-axis refers to the value of $N$, where we tried 5, 10, 15, 20, 25, 30, 35 and 40, though 5 and 10 are the most common length of item recommendations. They y-axis refers to the improvement ratio on F$_1$ or NDCG over the PR approach (i.e., the original Pareto ranking method) by using KD and the hybrid ranking methods. In order to present clearer visualizations, we only show the top two optimal hybrid rankings in Figure~\ref{fig:rst2}. Namely, they are the KD method using GD and PG as subsorting in the OpenTable data, and they are KD rankings using AR and MR as subsorting in the Yahoo!Movie data. In Figure~\ref{fig:rst2}, we can simply consider KD as a baseline, since it is a relaxed Pareto ranking method without subsorting, and observe whether the hybrid ranking methods can outperform the KD approach.

Based on the results on the OpenTable data, we can observe that the hybrid multi-criteria rankings can outperform the KD approach in both F$_1$ and NDCG. Particularly, PG is the overall winner as subsorting the OpenTable data, especially when $N$ is larger than 15. In the Yahoo!Movie data, using MR as subsorting can outperform the KD approach from the perspective of the F$_1$ metric. However, the hybrid ranking by using MR as subsorting can only outperform KD in NDCG when $N$ is no larger than 15. The hybrid ranking by using MR as subsorting can still be considered as an overall winner in the Yahoo!Movie data, since the NDCGs were not dropped significantly when $N$ is larger than 15. In a short summary, we can conclude that the hybrid ranking method by using subsorting can further offer improvements. However, there are no unique optimal method for subsorting. PG is the best for the OpenTable data, while MR is the best for the Yahoo!Movie data.

We can also discover more insights about these multi-criteria ranking methods from the results shown in Figure~\ref{fig:rst2}. We can observe that the improvement ratio goes down when we increase the value of $N$ in top-$N$ recommendations. The item recommendations are usually delivered as a short list and shown on the Internet applications, e.g., a widget on an e-commerce website, social media or online streaming. 5 and 10 are the most common choices for the value of $N$, though a longer list of recommendations may be required in specific domains, such as news or music recommendations. Based on the observations in Figure~\ref{fig:rst2}, on one hand, the improvement ratio is still positive, even if $N$ was increased to 40. On the other hand, KD and the hybrid ranking methods can offer significant improvements when $N$ is small, such as $N$ equals to 5 or 10 which are common options in recommender systems. However, the downtrend in Figure~\ref{fig:rst2} also reveals the drawbacks or weaknesses of the ranking methods based on dominance relation. This downtrend, most likely, is caused by the Pareto ranking or relaxed Pareto ranking, since they are the major sorting in hybrid ranking methods. The dominance relation may be helpful to help identify the most preferred items by users. However, the accuracy may go down when the length of preferred list is longer.

\section{Conclusions \& Future Work}
\noindent
In this paper, we proposed and examined hybrid multi-criteria ranking methods, where we utilize a Pareto ranking or relaxed Pareto ranking as the major sorting approach, and employ another preference ordering technique as subsorting in order to differentiate the ranking scores in the outcomes. Our experimental results over the OpenTable and Yahoo!Movie data sets can demonstrate the effectiveness of the proposed hybrid ranking methods. However, we also identify a new issue that the accuracy may go down if a longer list of item recommendations is required. We are planning to dig this issue and alleviate this in our future work.

\bibliographystyle{named}
\bibliography{sample-base}

\end{document}